\documentclass[12pt]{iopart}

\usepackage{latexsym}
\usepackage{graphicx}
\usepackage[activeacute,english]{babel}                 
\usepackage{bm}

\newcommand{\mean}[1]{\left \langle #1 \right \rangle}
\newcommand{\pd}[2]{\frac{\partial #1}{\partial #2}}

\newcommand{\ket}[1] {\left | \left. #1 \right \rangle \right.}
\newcommand{\bra}[1] {\left \langle \left. #1 \right | \right.}

\begin{document}

\title{Optimal protocols for Hamiltonian and Schr\"odinger dynamics}
  
\author{Tim Schmiedl, Eckhard Dieterich, Peter-Simon Dieterich, and Udo Seifert}
\address{{II.} Institut f\"{u}r Theoretische Physik, Universit\"{a}t Stuttgart, 70550 Stuttgart, Germany} 
\ead{useifert@theo2.physik.uni-stuttgart.de}
\date{\today}

\begin{abstract}
For systems in an externally controllable time-dependent potential, the optimal protocol minimizes the mean work spent in a finite-time transition between given initial and final values of a control parameter. For an initially thermalized ensemble, we consider both Hamiltonian evolution for classical systems and Schr\"odinger evolution for quantum systems. In both cases, we show that for harmonic potentials, the optimal work is given by the adiabatic work even in the limit of short transition times. This result is counter-intuitive because the adiabatic work is substantially smaller than the work for an instantaneous jump. We also perform numerical calculations of the optimal protocol for Hamiltonian dynamics in an anharmonic quartic potential. 
For a two-level spin system, we give examples where the adiabatic work can be reached in either a finite or an arbitrarily short transition time depending on the allowed parameter space.
\end{abstract}


\maketitle
\section{Introduction}

Recently, there has been considerable progress in the thermodynamic description of small (bio-) systems \cite{bust05, seif07} which are prone to thermal fluctuations and typically driven far out of equilibrium. It has been shown that thermodynamic quantities obey various exact work and fluctuation theorems \cite{evan93, gall95, jarz97,croo00,seif05a}. The Jarzynski relation connects the equilibrium free energy difference with an average of work values obtained from a series of nonequilibrium transitions \cite{jarz97, jarz97a}. 
While biomolecules are typically described by a Langevin equation where effects of the thermal bath are included in friction and fluctuation terms, the Jarzynski relation has originally been derived in the framework of purely Hamiltonian dynamics with initial canonical equilibrium distribution \cite{jarz97}. This situation corresponds to an experiment where the heat bath of temperature $T$ is decoupled from a small Hamiltonian system at time $\tau = 0$. During a time intevall $\tau \in [0,t]$ a control parameter $\lambda$ is varied time-dependently. Such dynamics has been used in various generalizations of the Jarzynski relation \cite{jarz00, kawa07} and for the optimization of the free energy reconstruction from nonequilibrium work data \cite{sun03, ober05, vaik08}. 

While this approach is elegant from a theoretical point of view, it is important to note that such a dynamics has quite different properties than a system subject to a permanent heat bath modelled {\sl e. g.} by a Langevin equation. The most im\-por\-tant difference concerns the quasistatic work which is achieved for infinitesimally slow transitions $t \to \infty$, independent of the detailed shape of the protocol for the control parameter $\lambda$. For a system coupled permanently to a heat bath conserving the canonical distribution for fixed control parameter, this quasistatic work is equivalent to the free energy difference. This property gives rise to the free energy calculation method termed thermodynamic integration where the free energy is approximated by the work for a very slow transition. For purely Hamiltonian dynamics, however, the quasistatic work (which then is called adiabatic work) can exceed the free energy difference \cite{ober05}. Still, the Jarzynski relation is valid and can be used to estimate free energy differences. It can be shown that under reasonable assumptions (mainly ergodicity), both the work is a monotonically decreasing function of the transition time $t$ for a given protocol shape and the adiabatic work is independent of the shape of the protocol \cite{alla07}.  

There has been considerable interest in deriving fluctuation theorems and analogues of the Jarzynski relation also for systems in the quantum domain \cite{muka03, dero04, enge07, talk07, croo08, espo08}.
%It is not obvious how to transfer results for classical Langevin dynamics to the quantum domain.  Moreover, 
Here, it is more subtle than in the classical case to define thermodynamic quantities such as work on a single trajectory \cite{enge07,talk07,croo08}. Apart from recent progresses for open quantum systems \cite{croo08}, most studies \cite{muka03, teif07, deff08, talk08} consider the case where the energy is measured at beginning and end, defining the work as the difference between final and initial energy. 
These studies rely on Schr\"odinger dynamics which has similar properties as Hamiltonian dynamics. In particular, the quasistatic work then is also different from the free energy difference.  

In this paper, we study optimal protocols of the control parameter $\lambda(\tau)$ leading to a minimal mean work for a finite given transition time $t$. This question has been tackled for Langevin dynamics in harmonic potentials \cite{schm07, gome08} with the result of unexpected singularities such as jumps in the overdamped case and even delta-type singularities in the underdamped case. For purely Hamiltonian or Schr\"odinger dynamics in harmonic potentials, we will show that the optimal protocol is degenerate and that the optimal work is equal to the adiabatic work. This result implies that the adiabatic work, which is achieved for an infinitely long transition time for any protocol, can be reached in an arbitrarily short time by using an appropriate protocol for the control parameter. This surprising result shows that purely Hamiltonian dynamics can beat Langevin evolution since for the latter short transition times, $t \to 0$, yield the work for an instantaneous jump. In the totally different context of collective escape over a barrier, it has been shown that Hamiltonian dynamics also beats Langevin dynamics in the sense that it leads to a faster escape \cite{henn07}. We will also calculate the optimal work numerically for an anharmonic potential. However, it is difficult to decide from the numerical data, whether the adiabatic work can be achieved in a finite transition time also for such an anharmonic potential.

\section{Hamiltonian dynamics}

In this section, we will study minimum work protocols for Hamiltonian dynamics in three different types of time-dependent potentials: (i) a moving harmonic potential, (ii) a harmonic potential with time-dependent stiffness and (iii) an anharmonic quartic potential. 

\subsection{Moving harmonic potential}
\label{Sect_Ham_mov_pot}
We consider a particle of mass $m$
subject to a moving harmonic potential 
\begin{equation}
V(x,\tau)=\frac{k}{2}(x-\lambda(\tau))^2,
\end{equation}
where $k$ is the (constant) stiffness of the potential. The minimum of the potential $\lambda(\tau)$ is changed time-dependently from an initial position $\lambda_i = 0$ to a final position $\lambda_f$. The Hamiltonian dynamics of position and momentum of the particle at time $\tau$ then is governed by the equations of motion
\begin{eqnarray}
\dot x &=& p / m , \nonumber \\
\dot p &=& -V'(x,\tau) , 
\label{eq:Ham_dyn}
\end{eqnarray}
where the prime denotes the derivative with respect to $x$ and the dot the derivative with respect to the time $\tau$. The particle initially is in thermal equilibrium with probability density
\begin{equation}
\rho(x_0,p_0) = \mathcal{N} \exp [-\beta V(x_0,0)] \exp[-\beta p_0^2 / (2m)]
\label{eq:Ham_rho}
\end{equation}
for the initial position $x_0$ and the initial momentum $p_0$ with normalization constant $\mathcal{N}$ and inverse temperature $\beta \equiv 1 / T$ with Boltzmann's constant $k_B \equiv 1$.

Since there is no heat transfer during the (purely Hamiltonian) transition, the work is given by the change of internal energy. Averaging over all initial conditions yields the mean work
\begin{eqnarray}
W &=& \left [\frac {\mean{p^2}} {2 m } + \mean{V} \right]_0^t \nonumber \\ &=&  \left [\frac {\mean{p^2}} {2 m } + \frac{k}{2} \left (\mean{x^2} - 2 \lambda \mean{x} + \lambda^2 \right )  \right]_0^t
\label{eq:Ham_W_mov}
\end{eqnarray}
where $\mean{ \cdot }$ denotes the average over the initial distribution of position and momentum (\ref{eq:Ham_rho}).
It can easily be shown that the variances $\mean{x^2} - \mean{x}^2$ and $\mean{p^2} - \mean{p}^2$ are time independent for any protocol $\lambda(\tau)$. Thus, the work  (\ref{eq:Ham_W_mov}) can be written as
\begin{eqnarray}
W &=& \left [\frac {\mean{p}^2} {2 m } + \frac{k}{2} \left (\mean{x}^2 - 2 \lambda \mean{x} + \lambda^2 \right )  \right]_0^t \nonumber \\
 &=& \left [\frac {\mean{p}^2} {2 m } + \frac{k}{2} \left (\mean{x}-\lambda \right )^2  \right]_0^t
\end{eqnarray}
which becomes zero, \textit{i.e.}, minimal for
\begin{eqnarray}
\mean{p(t)} &=& 0 ,  \nonumber \\
\mean{x(t)} &=& \lambda_f .
\label{eq:Ham_mov_cond}
\end{eqnarray}
The evolution equations for these mean values are given by
\begin{eqnarray}
\frac {\rm{d}} {\rm{d}\tau} \mean{x} &=& \mean{p} / m ,  \nonumber \\
\frac {\rm{d}} {\rm{d}\tau} \mean p  &=& - k (\mean x - \lambda)
\label{Eq:ham:mean}
\end{eqnarray}
with initial values $\mean{x(0)} = 0$ and $\mean{p(0)} = 0$.
With the freedom to choose the continuous function $\lambda(\tau)$ which corresponds to an infinite number of free parameters, it is quite obvious that the two final conditions can easily be met. In fact,  the optimal protocol is highly degenerate. 

For a specific choice, taking, {\sl e. g.}, a third order polynomial 
\begin{equation}
\lambda(\tau) = a (\tau/t) + b (\tau/t)^2 + (\lambda_f - a - b) (\tau / t)^3
\label{eq:poly_prot}
\end{equation}
with two free parameters $a,b$ suffices to meet the optimality conditions. The solution of the evolution equations (\ref{Eq:ham:mean}) can be solved analytically and, inserted into equation (\ref{eq:Ham_mov_cond}),  leads to
\begin{eqnarray}
a = \frac{12 \lambda_f \sin \left({\omega t}/{2}\right)-6 \omega t \lambda_f \cos
   \left({\omega t}/{2}\right)}{\left(12
   -\omega^2 t^2\right) \sin \left({\omega t}/{2
   }\right)-6 \omega t \cos
   \left({\omega t}/{2}\right)} ,  \nonumber \\
b = \frac{- 3 \omega ^2 t^2 \lambda_f \sin
   \left({\omega t}/{2}\right)}{\left(12
   -\omega ^2 t^2\right) \sin \left({\omega t}/{2
   }\right)-6 \omega t \cos \left( {\omega t} / 2 \right)}
\end{eqnarray}
with $\omega \equiv \sqrt{k / m}$.
For short transitions times $t \ll 1$, the coefficients diverge as $a \sim 1 / t^2$ and $b \sim - 1 / t^2$.
  
As a quite different choice, both conditions can also be met by a linear protocol with delta singularities at the boundaries, in analogy to the underdamped stochastic dynamics case \cite{gome08}. Indeed, it can easily be verified that the protocol
\begin{equation}
\lambda^*(\tau) = \lambda_f \frac{\tau}{t} +  {\frac{m \lambda_f}{k t}} \left [ \delta(\tau) -   \delta(\tau-t) \right ]
\end{equation}
fulfills the conditions (\ref{eq:Ham_mov_cond}). This expression could also have been obtained as the limit of the underdamped optimal protocol for a vanishing friction coefficient $\gamma \to 0$. 

Thus, for any given total transition time $t$ the minimal mean work is given by
\begin{equation}
W^* = 0 .
\end{equation}
For infinitely long transition times, quite generally, the work is given by the adiabatic work. For the studied model system, we have $W^{\rm{ad}} = 0$. As shown above, this lower bound on the work can even be achieved for arbitrarily short transition time. This surprising result shows that purely Hamiltonian dynamics can beat Langevin evolution where short transition times $t \to 0$ yield the work for an instantaneous jump $W^{\rm{jp}} = k \lambda_f^2 / 2$. 

The optimal protocols show a distinct positive peak in the first part of the transition and a distinct negative peak in the second part of the transition. This peak structure can be understood intuitively in complete analogy to the underdamped dynamics case \cite{gome08}. The first peak accelerates the particle and leads to a finite mean velocity which is necessary to reach a final mean particle position at the potential minimum. The second negative peak decelerates the particle, thus recovering the invested work from the kinetic energy of the particle.

\subsection{Harmonic potential with time dependent stiffness}
\begin{figure}
\begin{center}
 \includegraphics[width = 0.6 \linewidth]{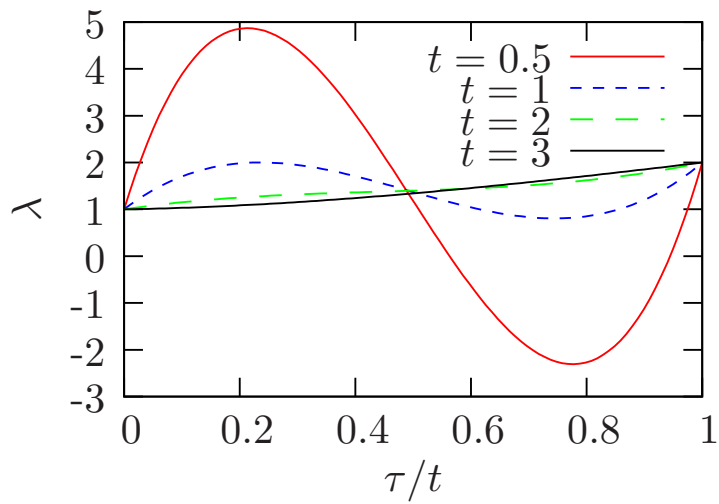}
\caption{Optimal third order polynomial protocols $\lambda_p^*(\tau/t)$ for the harmonic potential with time-dependent stiffness, see equation (\ref{pot_stiff}), with $m=1, T=1, k=1, \lambda_f = 2, \lambda_i = 1$ for different transition times $t$. For short transition times, the optimal protocol shows a pronounced positive peak in the first part of the transition and a pronounced negative peak in the second part of the process, thereby regaining the invested work.
\label{fig:Ham_stiff_prot}}
\end{center}
\end{figure}

We consider the time dependent potential
\begin{equation}
V(x,\tau)=\frac{\lambda(\tau)}{2}x^2,
\label{pot_stiff}
\end{equation}
where the stiffness $\lambda(\tau)$ is changed from an initial value $\lambda(0) = \lambda_i$ to a final value $\lambda(t) = \lambda_f$ in a finite time $t$. The work exerted on the system during the finite time transition is given by the change in internal energy
\begin{equation}
W = \left [\frac {\mean{p^2}} {2 m } + \mean{V} \right]_0^t =  \left [\frac {\mean{p^2}} {2 m } + \frac \lambda 2 \mean{x^2} \right]_0^t.
\end{equation}
The dynamics  (\ref{eq:Ham_dyn}) leads to coupled evolution equations for the mean squared position
$w \equiv \langle x^2 \rangle$ and the mean squared momentum $z \equiv \langle p^2 \rangle$
\begin{eqnarray} 
\dot{z} &= &- m \lambda \dot w  , \label{eq:Ham_dotz} \\
\ddot{w} & = & \frac {2 z} {m^2} -\frac 2 m \lambda w \label{eq:Ham_ddotw}  .
\end{eqnarray}
Multiplication of equation (\ref{eq:Ham_ddotw}) with a factor $\dot w$ and insertion of equation (\ref{eq:Ham_dotz}) leads to the relation
\begin{equation}
\ddot w \dot w = \frac {2} {m^2} \left ( z \dot w + w \dot z \right) 
\end{equation}
which, using the equilibrium initial conditions $w(0) = T / \lambda_i$, $\dot w(0) = 0$, and $z(0) = m T$ can be integrated to yield
\begin{equation}
zw = \frac {m^2} {4} \dot w^2 + \frac{m T^2} {\lambda_i}.
\end{equation}
At the final time $t$, minimization requires $\dot w(t) = 0$ which is equivalent to the final condition 
\begin{equation}
z^*(t) = \frac{m T^2} {\lambda_i w(t)}.
\label{eq:Ham_z_opt}
\end{equation}
The (mean) total work thus becomes
\begin{equation}
W = \frac 1 2 \left [\lambda_f w(t) - T + \frac {T^2} {w(t) \lambda_i} - T \right ].
\end{equation}
Optimizing this expression with respect to $w(t)$ leads to 
\begin{equation}
w^*(t) = \frac {T} {\sqrt{\lambda_i \lambda_f}} 
\label{eq:Ham_w_opt}
\end{equation}
which yields the optimal work
\begin{equation}
W^* = T \left (\sqrt{\frac {\lambda_f} {\lambda_i} } - 1 \right ).
\label{eq:meanW_Ham}
\end{equation}
Similar to the first case study, the optimal protocol is highly degenerate. 
For a third order polynomial, the equations of motion cannot be solved analytically. However, as expected, the two parameters can be chosen such that the boundary conditions (\ref{eq:Ham_z_opt}) and (\ref{eq:Ham_w_opt}) are fulfilled. Optimal third order polynomial protocols $\lambda_p^*(\tau/t)$ are shown in Fig. \ref{fig:Ham_stiff_prot} for different transition times $t$. A short transition time requires pronounced peaks with increasing height for decreasing transition times. These peaks serve to accelerate the Brownian particle in the first part of the transition and to deccelerate the particle in the second part of the transition.

Another convenient choice is a linear protocol with superimposed delta peak singularities at beginning and end. The initial singularity allows for setting the initial mean squared momentum instantaneously to a value $z(0+) \neq z(0) = m T$ which can be tuned to meet the boundary condition on $w(t) = w^*(t)$. The second delta singularity at the final time $t$ allows for setting  $\dot w(t) = 0$.

Note that, similar to the first case study, the minimal work (\ref{eq:meanW_Ham}) does not depend on the transition time $t$.  Naively, in the limit $t \to 0$, one would expect the result obtained for an instantaneous jump of the protocol $W^{\rm{jp}} = (\lambda_f - \lambda_i) / 2$ which is substantially larger than the optimal work. However, the possibility to use a protocol with increasing absolute values of $\lambda$ (and thus increasing forces) for decreasing transition times $t$ leads to the singular limit at $t \to 0$. For any fixed shape of the protocol $\lambda(\tau / t)$, the work approaches $W^{\rm{jp}}$ with $t \to 0$.

\subsection{Anharmonic quartic potential}

In both harmonic case studies, the optimal work $W^*$ is given by the adiabatic work $W^{\rm{ad}}$, independent of the total transition time $t$. The optimal protocols $\lambda^*(\tau)$ are highly degenerate. It is interesting to see whether these (unexpected) features persist for anharmonic potentials where the optimality conditions cannot be cast into only two conditions for the moments of the probability distribution. However, it is computationally quite difficult to determine optimal protocols in the anharmonic case. For a case study, we use the quartic potential
\begin{equation}
V(x,\lambda) = \frac 1 4 \lambda x^4
\label{Ham_anharm_pot}
\end{equation}
with $T=1$, $\lambda_i = 1$, and $\lambda_f=2$ together with a Fourier ansatz for the optimal protocol
\begin{equation}
 \lambda^*(\tau) = 1 + \frac \tau {t} + a_0 \sin(\pi \tau / t) + \sum_{k=1}^n a_k \sin(2 k \pi \tau / t)
\label{eq:Ham_Fourier_ansatz}
\end{equation}
where we omit higher odd frequencies because their symmetry is not suitable to produce the peak structure found in the two harmonic cases.
We next optimize the parameters $a_k$ for a minimal mean work which we calculate as the average of work values from $25000$ randomly sampled initial values for $x_0$ and $p_0$. For the minimization, we use a standard Mathematica algorithm. Note that the optimal work values obtained in this numerical procedure are upper bounds for the true optimal work since we have used a finite number of parameters. 

The adiabatic work can be calculated using the microcanonical distribution for given energy $E$ and given $\lambda$. The change of internal energy then is given by
\begin{equation}
\frac {dE} {d\lambda} = \mean{\pd {V} \lambda}_{\rm{micro}}  \equiv \frac 1 Z \int dx \frac {1} {\sqrt{2m \left(E - V(x,\lambda)\right)}} \pd {V(x, \lambda)} {\lambda}
\label{eq:Ham_dgl_qs}
\end{equation}
with the microcanonical partition function
\begin{equation}
Z \equiv \int dx \frac {1} {\sqrt{2m \left ( E - V(x,\lambda)\right)}}.
\end{equation}
The evolution equation (\ref{eq:Ham_dgl_qs}) is an ordinary differential equation for $E(\lambda)$.
For the potential $V(x, \lambda) = \lambda x^4 / 4$, we obtain 
\begin{equation}
\frac {dE} {d\lambda} = \frac{E} {3 \lambda}
\end{equation}
which has the solution $E = E_0 \lambda^{1/3}$ with initial energy $E_0$.
Averaging over initial conditions $x_0, p_0$ according to the appropriate initial Boltzmann distribution (\ref{eq:Ham_rho}) then yields
\begin{equation}
W^{\rm{ad}} = 3 \frac {(\sqrt[3] {2} - 1) \Gamma (5/4) } {\Gamma (1/4)} \simeq 0.1949 .
\end{equation}

\begin{figure*}
\includegraphics[width = 0.49 \linewidth]{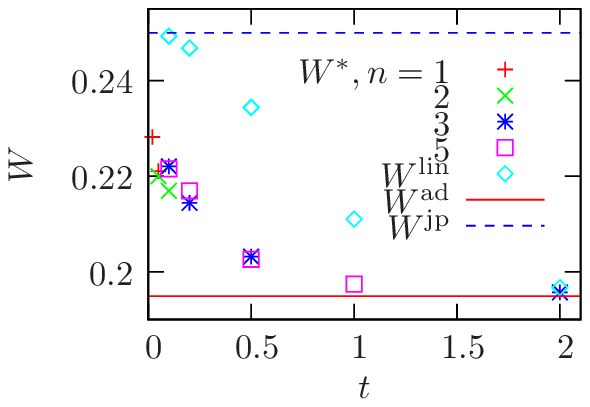}
\includegraphics[width = 0.49 \linewidth]{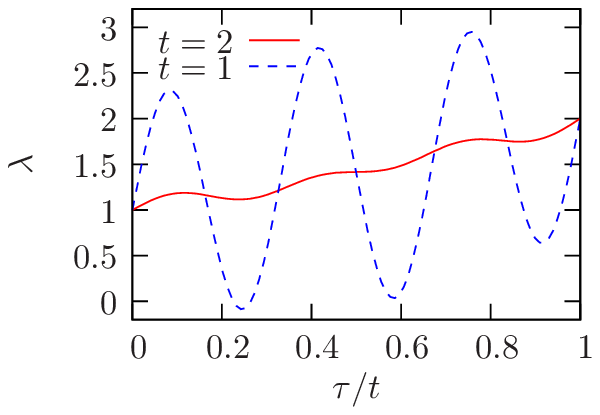}
\caption{Optimization results for the anharmonic potential (\ref{Ham_anharm_pot}). (a) Optimal work  from the Fourier ansatz (\ref{eq:Ham_Fourier_ansatz}) as a function of the transition time $t$ for different number of Fourier modes $n$ compared to the work  $W^{\rm{lin}}$ obtained for the linear protocol $\lambda^{\rm{lin}} \equiv 1 + \tau / t$ and the adiabatic work $W^{\rm{ad}} \simeq 0.1949$. (b) Optimal protocols as obtained from the Fourier ansatz (\ref{eq:Ham_Fourier_ansatz}) with $n = 3$ for different transition times $t$.
\label{fig:Ham_x4_W}}
\end{figure*}

In Fig. \ref{fig:Ham_x4_W}a, we compare optimal work values with the adiabatic work. 
%It is hard to decide from the numerics whether this adiabatic work can be reached in a finite time.
  Even though it looks like convergence may be reached already by
the small number of Fourier modes, we cannot decide
this question conclusively. Particularly for very small transition times,
optimal protocols feature extreme protocol values which may require a large
number of Fourier modes. Moreover, we cannot exclude the possibility that
the minimum search is trapped in a local minimum preventing the convergence
towards the global minimum. Therefore, we cannot exclude the possibility that 
the adiabatic work can be reached in a finite time also in this case study.

 More likely,
however, is the following scenario:
The optimal work is discontinuous at $t=0$. For any transition time,
work values well below the work for an instantaneous jump $W^{\rm jp} = 0.25$ but above
the adiabatic work can be obtained.
Clearly, even for very fast transitions compared to the characteristic oscillation time $t_{\rm{char}} \simeq 3$, the optimal work is well below the work for a linear protocol which has the limit $W^{\rm jp} = 0.25$ for $t \to 0$. The optimal protocol again shows pronounced peaks for short transition times $t$, see Fig. \ref{fig:Ham_x4_W}b. Since we use an ansatz with a finite number of parameters, it is impossible to decide whether the optimal protocol is degenerate.

\begin{figure*}
\begin{center}
\includegraphics[width = 0.48 \linewidth] {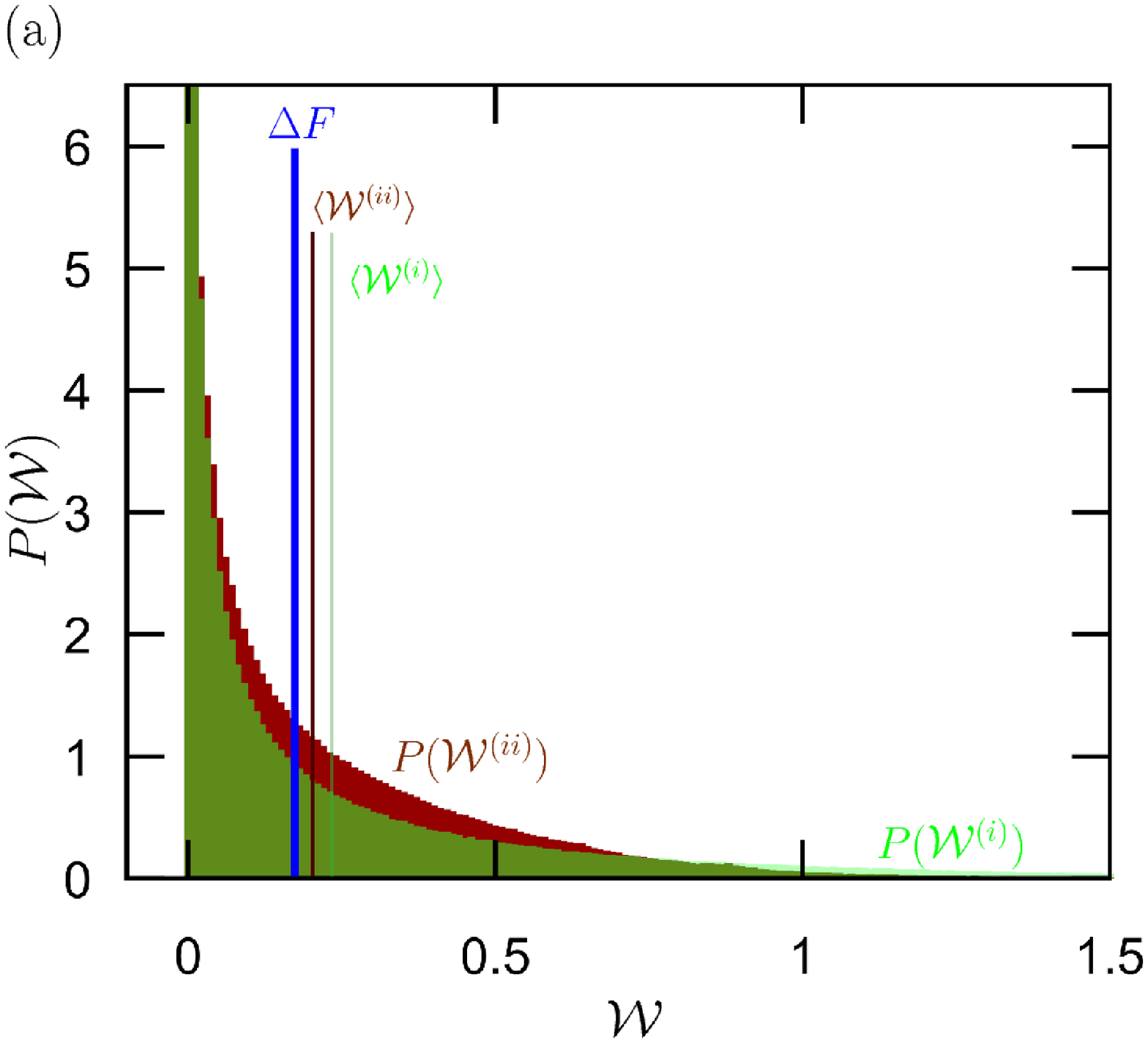}
\includegraphics[width=0.48 \linewidth]{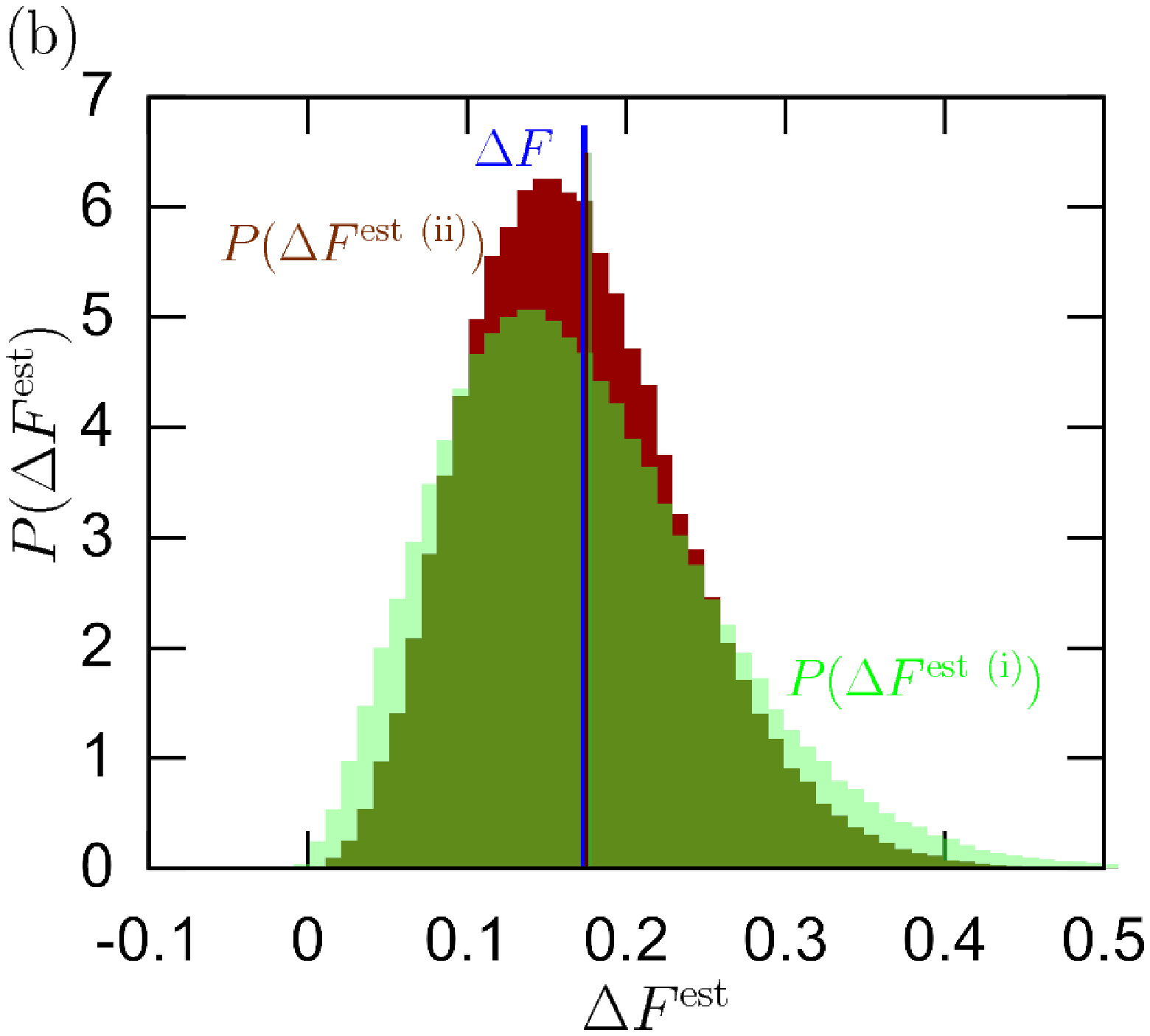}
\caption{Comparison of free energy estimates for the anharmonic potential (\ref{Ham_anharm_pot}) for (i) a linear protocol and (ii) the optimized Fourier protocol ($n=3$) for  $t = 0.5$. The data were obtained from Langevin simulation of $2 \cdot 10^6$ trajectories for each protocol with $m = 1$, $T=1$, and $\lambda_i = 1$, $\lambda_f = 2$.  (a) Distribution $P(\mathcal{W})$ of work values $\mathcal{W}$ for the two protocols. The columns show the free energy difference $\Delta F \simeq 0.1733$ and the mean work values $\mean{\mathcal{W}^{(i)}} \simeq 0.2344 (\pm 0.0003)$, $\mean{\mathcal{W}^{(ii)}} \simeq 0.2025 (\pm 0.0002)$. (b) Histogram of $2 \cdot 10^5$ Jarzynski estimates $\Delta F^{\rm{est}} \equiv - (1 / \beta) \ln \left [\sum_{i=1}^N \exp (-\beta \mathcal{W}_i) / N \right ]$ for the free energy difference obtained from $N = 10$ single trajectory work values $\mathcal{W}_i$ each. The mean squared error (MSE) of these estimates consists of two parts: the systematic error (bias) $B = \mean{\Delta F^{\rm{est}}} - \Delta F$ and  the statistical error $\sigma = \sqrt{\rm{Var} (\Delta F^{\rm{est}} )}$. The columns show the free energy difference and the mean value of the estimates obtained from the two protocols. Since the bias ($B^{(i)} = 0.0038 (\pm 0.0002)$, $B^{(ii)} = 0.0018 (\pm 0.0002)$) can be neglected for both protocols, the MSE is dominated by the statistical error ($\sigma^{(i)} = 0.0843$, $\sigma^{(ii)} = 0.0665$) which is larger by  $27 \%$ for protocol (i).
\label{fig_dF}}
\end{center}
\end{figure*}

 Even though we could not decide whether the adiabatic work can be reached in a finite time, we can show that free energy differences as determined via the Jarzynski relation indeed improve for an optimal protocol compared to a linear protocol, see Fig. \ref{fig_dF}  and its caption for details.

\section{Quantum Systems}
In this section, we investigate optimal protocols for quantum system. 
Fortunately, for harmonic systems which are comparable to the first two case studies from the previous sections, work distributions $p(W)$ have recently been calculated for Schr\"odinger dynamics \cite{deff08,talk08}. By using their results, it is easy to derive optimal protocols since in both cases, the work distributions (and therefore also the mean work) depend on a single ``rapidity'' parameter which vanishes for quasistatic transitions. In the last subsection, we study the two level system of a single spin in a time dependent magnetic field. If we impose that the absolute value of the magnetic field stays fixed, the adiabatic work can only be reached in a finite (non-zero) time.

\subsection{Moving harmonic potential}
We consider the quantum analogue of a moving harmonic potential 
\begin{equation}
V(x,\tau)=\frac{m \omega^2}{2}(x-\lambda(\tau))^2
\end{equation}
with Hamiltonian
\begin{equation}
H = \hbar \omega \left[ a^\dag a - \sqrt{\frac {m \omega} {2 \hbar}} \lambda(\tau) \left (a +a^\dag \right) \right] + \frac {m \omega^2} {2} \lambda(\tau)^2 
\end{equation}
with creation and annihilation operators $a^\dag$ and $a$, Planck constant $\hbar$, frequency $\omega$ and a time dependent minimum of the harmonic potential $\lambda(\tau)$ which is changed from an initial value $\lambda(0) = 0$ to a final value $\lambda(t) = \lambda_f$ during a given finite time $t$.  It has recently been noted that the mean work is given exactly by the classical expression for this case study \cite{camp08}. Thus, the optimal protocols as discussed in Sect. \ref{Sect_Ham_mov_pot} also apply to the quantum case. However, the work distributions are different for classical and quantum systems. It has been shown  \cite{talk08, talkner_footnote} that the work distribution $p(W)$ for an initially thermalized ensemble depends only on the absolute value $\left | z \right |$ of a ``rapidity'' parameter 
\begin{equation}
z \equiv \sqrt \frac  {m \omega} {2 \hbar}  \int_0^t d\tau \dot \lambda(\tau) e^{i \omega \tau}
\end{equation}
which vanishes for adiabatic transitions with $t \to \infty$. However, even for transitions in a finite time $t$, there exist appropriate protocols $\lambda(\tau)$ which lead to a vanishing ``rapidity parameter'' $z$ and thus to a vanishing mean work $W^* = W^{\rm{ad}} = 0$. 
In addition to the optimal protocols from Sect. \ref{Sect_Ham_mov_pot}, it also suffices to consider the simple trigonometric protocol
\begin{equation}
\lambda^*_n(\tau) = {\lambda_f} \left[ \frac{\tau }{t}+\frac{\left(4 \pi^2 n^2-t^2 \omega ^2\right) \sin
   \left(\frac{2 n \pi  \tau }{t}\right)}{2 n \pi t^2 \omega ^2} \right ]
\end{equation}
which can be easily shown to yield $z = 0$ for any integer $n$. 
%{\bf Similar protocols have been shown to yield zero mean work for a finite time in Ref. \cite{camp08}. However, it has not been noticed there that the adiabatic work can be achieved for any arbitrary short transition time.} 
%Thus, the adiabatic work can also be achieved in an arbitrarily short transition time. 
Again, the optimal protocol is highly degenerate. This degeneracy arises because the continuous protocol $\lambda(\tau)$ with an infinite number of parameters has to satisfy only two boundary conditions $\lambda(0) = 0$, $\lambda(t) = \lambda_f$ and the adiabaticity condition $z=0$.

\subsection{Harmonic potential with time dependent frequency}
A similar argument can be applied to the quantum analogue of the harmonic potential with time-dependent stiffness $m \lambda(\tau)$ with Hamiltonian
\begin{equation}
H = \frac {p^2}{2 m} + \frac m 2 \lambda(\tau) x^2.
\end{equation}
The control parameter $\lambda$ is changed time-dependently from an initial value $\lambda(0) = \lambda_i$ to a final value $\lambda(t) = \lambda_f$ during a finite transition time $t$, which corresponds to a time-dependent oscillator frequency $\omega(\tau) \equiv \sqrt{\lambda(\tau)}$. It has been shown \cite{deff08} that the work distribution and therefore also the mean work only depends on a single parameter
\begin{eqnarray}
Q = \frac 1 {2 \sqrt{\lambda_i \lambda_f}} && \left [  \lambda_i \left(\lambda_f X(t)^2 + \dot X(t)^2\right) \right. \nonumber \\ 
 &&+ \left. \left(\lambda_f Y(t)^2 + \dot Y(t)^2\right) \right ] - 1
\end{eqnarray}
where $X(t)$ and $Y(t)$ are solutions for the classical harmonic oscillator equation 
\begin{equation}
\ddot X(\tau) + \lambda(\tau) X(\tau) = 0
\end{equation}
with initial conditions 
\begin{equation}
X(0) = 0~,~~\dot X(0) = 1~,~~Y(0) = 1~,~~\dot Y(0) = 0 .
\label{eq:Ham_QM_stiff_cond}
\end{equation}
The dynamics of $X(\tau)$ and $Y(\tau)$ is constrained by their time evolution with common protocol $\lambda(\tau)$ which yields the relation
\begin{equation}
\ddot X Y = \ddot Y X.
\end{equation}
Addition of the term $\dot X \dot Y$ on both sides and subsequent integration leads to
\begin{equation}
\dot X Y = X \dot Y + 1
\label{eq:Ham_constr}
\end{equation}
where the integration constant is determined by the initial conditions (\ref{eq:Ham_QM_stiff_cond}).
Using standard techniques, we perform a minimization of $Q$ under the constraint (\ref{eq:Ham_constr}) with respect to $X(t)$, $\dot X(t)$, $Y(t)$, and $\dot Y(t)$. The optimality conditions then are given by
\begin{eqnarray}
X^*(t) &=& \frac 1 {\sqrt{\lambda_i \lambda_f}} \sqrt{\sqrt{\lambda_i \lambda_f} - \lambda_f Y^*(t)^2} \nonumber \\ 
\dot X^*(t) &=& \sqrt{\frac {\lambda_f}{\lambda_i}} Y^*(t)
\end{eqnarray}
where $Y^*(t)$ can be chosen arbitrarily. The minimal value of the parameter $Q$ then is $Q^* = 0$ and the optimal work is given by the adiabatic work which can be calculated \cite{deff08} as
\begin{equation}
W^* = W^{\rm{ad}} = \frac 1 2 \hbar (\omega_f -\omega_i) \coth(\beta \hbar \omega_i / 2)
\end{equation}
where $\omega_f \equiv \sqrt {\lambda_f}$ and $\omega_i \equiv \sqrt {\lambda_i}$. Again, only a finite number of optimality conditions have to be fulfilled while the continuous protocol $\lambda(\tau)$ corresponds to an infinite number of parameters. The optimal protocol $\lambda^*(\tau)$ thus is highly degenerate and the adiabatic work can be reached in any given (arbitrarily short) transition time.
 
\subsection{Two-level system}
We next study a simple two level system consisting of a single spin in a time dependently rotating magnetic field ${\bf B}(\tau) = B {\bf e} (\tau)$ where ${\bf e}(\tau)$ is a unit vector which is to be changed from an initial value ${\bf e}(0) =  \hat {\bm e}_z$ to a final value ${\bf e}(t) =  -\hat {\bm e}_z$ in a given finite transition time $t$, keeping the absolute value of the magnetic field $B$ fixed. Here $\hat {\bm e}_i$ denotes the unit vector in $i$-direction, $i \in \lbrace x,y,z \rbrace$.  The Hamiltonian of the system is given by 
\begin{equation}
H = - \hbar \gamma {\bf B} {\bm \sigma} / 2 
\end{equation}
with Pauli matrices ${\bm \sigma}$ and gyromagnetic ratio $\gamma$. In the basis of eigenvectors of ${ \sigma}_z$,  the initial canonical density matrix is given by
\begin{equation}
\rho(0) = \frac 1 Z \left (e^{\beta \hbar \gamma B /2} \ket{+} \bra{+}~~ +~~ e^{- \beta \hbar \gamma B /2} \ket{-} \bra{-} \right ),
\end{equation}
where $ \ket{+}$ is eigenvector of ${ \sigma}_z$ with eigenvalue $+1$, $ \ket{-}$ is eigenvector of ${ \sigma}_z$ with eigenvalue $-1$ and $Z$ is the canonical partition function. The work in such a finite time transition then is given by
\begin{equation}
W = \mean{E}_t - \mean{E}_0 = \tr\{H(t) \rho(t)\} - \tr \{H(0) \rho(0) \}
\end{equation}
where the time evolution of the density matrix $\rho(\tau)$ is given by the Liouville-von-Neumann equation
\begin{equation}
\partial_{\tau} \rho(\tau) = \frac {i} {\hbar} \left[\rho(\tau), H(\tau) \right ] .
\end{equation}

Quite generally, due to the adiabatic theorem in quantum mechanics, the occupation probabilities of the states do not change for quasistatic driving. The density matrix at the end of an adiabatic transition is therefore given by
\begin{equation}
\rho_1 \equiv \frac 1 Z \left ( e^{-\beta \hbar \gamma B /2} \ket{+} \bra{+}~~ +~~ e^{\beta \hbar \gamma B /2} \ket{-} \bra{-} \right).
\end{equation}
Thus, initial and final energy are equal $\tr\{H(0) \rho(0)\} = \tr \{H(t_c) \rho(t_c) \}$ and the adiabatic work is zero $W^{\rm{ad}} = 0$.

We next try to identify optimal protocols for such a transition. Generally, any pure spin state corresponds to a Bloch vector of unit length. Applying a magnetic field ${\bf B}$ leads to a precession of the Bloch vector around the axis of the magnetic field with angular frequency $\omega_L \equiv \gamma B$. Therefore, after an appropriate time $t_c = \pi / \omega_L$, the direction of a Bloch vector perpendicular to the magnetic field is reversed. During the adiabatic evolution from initial state $\rho_0$ to the final state $\rho_1$, the initial Bloch vectors in $\hat {\bm e}_z$ and $- \hat {\bm e}_z$ directions are also reversed. In order to reach the adiabatic work in a total transition time $t = t_c$, it therefore suffices to choose the protocol
\begin{equation}
{\bf e^*}(\tau) = \left \lbrace \begin{array} {l} \hat {\bm e}_z~~\mathrm{for}~~ \tau = 0 \\ \hat {\bm e}_x~~\mathrm{for}~~ 0 < \tau < t_c  \\ -\hat {\bm e}_z~\mathrm{for}~~ \tau = t_c \end{array} \right.
\end{equation}
which corresponds to an initial jump to a (constant) magnetic field perpendicular to the $z$-direction, \textit{i.e.} ${\bf B} = B \hat {\bm e}_x$, and a final jump to the magnetic field ${\bf B} = -B \hat {\bm e}_z$. At the end of the transition, the density matrix is given by
\begin{equation}
\rho(t_c) = \rho_1
\end{equation}
and the mean work thus is equal to the adiabatic work $W^* = W^{\rm{ad}} = 0$. The adiabatic work thus can be reached in the finite time $t_c$. In order to reach the adiabatic work for any larger transition time $t>t_c$, it suffices to choose the protocol ${\bf e^*}(\tau)$ for times $t \leq t_c$, keeping the magnetic field ${\bf B}(\tau > t_c) = -B \hat {\bm e}_z$ unchanged for all times $\tau>t_c$. For transition times $t<t_c$, the adiabatic work cannot be reached since there is no possibility to obtain the density matrix $\rho_1$ from the given dynamics in a time $t<t_c$.

However, if we had not constrained the magnetic field to have a given absolute value $B$ but allowed for any value of the magnetic field during the transition, we would have found that the adiabatic work can be reached in any transition time $t$. This is obvious because the spin precession frequency is proportional to $B$ and thus any precession frequency can be achieved in such a setup. For a given transition time $t$, the magnetic field then can be tuned to yield a precession frequency $\omega_L = \pi / t$ which leads to  $W^* = 0$.

More generally, if one allows any form of the Hamiltonian, it is obvious that a work value arbitrarily close to the adiabatic work can be reached in any given (short) time $t$ for the following reason. Assume that the work $W_1$ can be reached in a slow transition with dynamics $\Psi_1(\tau)$ for the time dependent Hamiltonian $H_1(\tau)$ with $t_1 \gg t$. The Hamiltonian $H(\tau) \equiv  ( t_1 / t )  H_1(t_1  \tau / t) $ then leads to an equivalent dynamics $\Psi(\tau) = \Psi_1(t_1  \tau / t)$ and therefore to an equivalent final mean internal energy at time $t$. Thus, the work $W_1$ (which is arbitrarily close to $W^{\rm{ad}}$ for long times $t_1$) can be reached in the (arbitrarily short) time $t$. Note, however, that this argument is no longer true if one allows only for any form of the potential $V(x,\tau)$ since the kinetic energy (and thus the dependence on the momentum $p$) in the Hamiltonian $H(\tau)$ is given and thus cannot be rescaled in this manner.

\section{Conclusion and Perspectives}
We have investigated optimal protocols for a minimal mean work for both Hamiltonian and Schr\"odinger dynamics with canonical initial conditions.  This extends our previous studies of optimal protocols for overdamped and underdamped Langevin dynamics \cite{schm07,gome08} where we found unexpected jumps in the overdamped case and even delta peaks in the underdamped case. 

For Hamiltonian dynamics and harmonic potentials, the present optimization shows two unexpected features  which do not occur for Langevin dynamics: (i) Independently of the transition time $t$, the optimal work $W^*$ is given by the adiabatic work obtained for an infinitely slow transition, and (ii) the optimal protocol $\lambda^*(\tau)$ is highly degenerate since only two conditions on the moments of the probability distribution have to be fulfilled. 

For Hamiltonian dynamics in an anharmonic potential, we performed numeric calculations of the optimal protocol in a restricted state space. We found that the optimal work is well below the work for an instantaneous jump even for very fast transitions and we thus conjecture that the optimal work is quite generally discontinuous at $t = 0$. However, we could not decide from our numerical data whether the adiabatic work can be obtained in an arbitrarily short time $t$ also for anharmonic potentials. In order to decide this question conclusively, extensive numerics is necessary which will be left for future research. 

For both harmonic and anharmonic potentials, values very close to the adiabatic work can be reached with transition times well below the typical oscillation times. These results suggest that the efficiency of free energy calculations for Hamiltonian dynamics via the Jarzynski relation can be significantly improved by using the optimal protocol. This insight is even more relevant since thermodynamic integration leads to a biased estimate because of the inequality of adiabatic work and free energy difference. 

Quite generally, the question whether the adiabatic work can be achieved in an arbitrarily short time will depend on the dependence of the potential $V(x,\lambda)$ on the protocol $\lambda(\tau)$.
In the harmonic case studies, any constraint on the maximal possible absolute value of $\lambda(\tau)$ for Hamiltonian dynamics would lead to optimal work values well above the adiabatic work for short transition times. Such a constraint could, {\it e.~g.}, be implemented by replacing $\lambda(\tau) \equiv a \tanh(\tilde \lambda(\tau))$, which yields values between $\lambda_{\rm{min}} = -a$ and $\lambda_{\rm{max}} = a$ for any value of $\tilde \lambda$. Consequently, the question arises whether the adiabatic work can be reached in an arbitrary short time using any type of intermediate potentials which does include  not only those compatible with a given functional form $V(x,\lambda)$. It is obvious that, if one allowed also for a time-dependent mass $m(\tau)$, the adiabatic work could be reached in any given time, in complete analogy to the discussion in the quantum case.

For quantum systems in an initially thermalized state, minimal work protocols show the same features as in the classical case for harmonic potentials. For a two level system which could serve as a paradigm for a finite quantum system, it depends on the allowed parameter space for the ``potential'', {\it i.~e.} the magnetic field, whether the adiabatic work can be reached in an arbitrarily short time. 
Generally, the adiabatic work can be reached in arbitrarily short time if any functional form of the time-dependent Hamiltonian $H(\tau)$ is allowed.

For most practical purposes, the range of the control parameter will be limited by the experimental conditions. For quantum systems, refined strategies for the optimal control of the time evolution of wave functions under such constraints have been developped \cite{shi88, kosl89, juds92, yan93, gord97}. These techniques may be generalized in order to specify minimum work protocols for quantum nonequilibrium transitions in complex systems with additional constraints on the possible values of the control parameter $\lambda(\tau)$.

\end{document}